\documentclass[fleqn]{2023SCGE}
\usepackage{siunitx}

\begin{document}

\ensubject{subject}

\ArticleType{Article}
\SpecialTopic{SPECIAL TOPIC: }
\Year{XXX}
\Month{XXX}
\Vol{XXX}
\No{XXX}
\DOI{XXX}
\ArtNo{XXX}
\ReceiveDate{XXX}
\AcceptDate{XXX}

\title{Quantum Clock Synchronization Network with Silicon-chip Dual-Pumped Entangled Photon Source}{Quantum Clock Synchronization Network with Silicon-chip Dual-Pumped Entangled Photon Source}
\author[1,2$^\dag$]{Jiaao Li}{}
\author[3,4$^\dag$]{Hui Han}{}
\author[1,2]{Xiaopeng Huang}{}
\author[5]{Bangying Tang}{}
\author[6]{Kai Guo}{}
\author[4,7]{Jinquan Huang}{}
\author[4,8]{\\Siyu Xiong}{}
\author[3]{Wanrong Yu}{}
\author[2]{Zhaojian Zhang}{}
\author[2]{Junbo Yang}{}
\author[4]{Bo Liu}{{liubo08@nudt.edu.cn}}
\author[2]{Huan Chen}{{chenhuan11@nudt.edu.cn}}
\author[1]{Zhenkun Lu}{{lzk06@sina.com}}

\AuthorMark{J. A. Li}

\AuthorCitation{J. A. Li, H. Han, X. P. Huang, et al}
\address[1]{College of Electronic Information, Guangxi Minzu University, Nanning 530006, China}
\address[2]{College of Science, National University of Defense Technology, Changsha 410073, China}
\address[3]{College of Computer, National University of Defense Technology, Changsha 410073, China}
\address[4]{College of Advanced Interdisciplinary Studies, National University of Defense Technology, Changsha 410073, China}
\address[5]{Strategic Assessments and Consultation Institute, Academy of Military Sciences, Beijing 100091, China}
\address[6]{Institute of Systems Engineering, Academy of Military Sciences, Beijing 100039, China}
\address[7]{School of Electronics and Communication Engineering, Shenzhen Campus of Sun Yat-sen University, Shenzhen, 518107, China}
\address[8]{School of Mathematical Sciences, Sichuan Normal University, Chengdu 610068, China}


\abstract{In this paper, we propose a quantum clock synchronization (QCS) network scheme with silicon-chip dual-pumped entangled photon source. This scheme couples two pump beams into the silicon-based waveguide, where degenerate and non-degenerate spontaneous four-wave mixing (SFWM) occurs, generating entanglement between one signal channel and three idler channels. The entangled photons are distributed to remote users through the wavelength division multiplexing strategy to construct an entanglement distribution network, and the round-trip QCS is adopted to realize a QCS network that can serve multiple users.   A proof-of-principle QCS network experiment is implemented among the server and multiple users (Alice, Bob, and Charlie) for 11.1 hours, where Alice and Charlie are \SI{10}{\kilo\meter} away from the server and Bob is \SI{25}{\kilo\meter} away from the server. 
The lowest time deviations (TDEV) between the server and each user (Alice, Bob, and Charlie) are \SI{1.57}{\pico\second}, \SI{0.82}{\pico\second} and \SI{2.57}{\pico\second} at the average time of \SI{8000}{\second}, \SI{8000}{\second} and \SI{800}{\second} respectively. 
The results show that the QCS network scheme with dual-pumped SFWM photon source proposed by us achieves high accuracy, and the channel resources used by $n$ users are reduced by about 30\% compared with other round-trip QCS schemes.}



\maketitle


\begin{multicols}{2}
\section{INTRODUCTION}\label{section1}
All along, high-precision clock synchronization has been an indispensable technology in numerous fields such as navigation \cite{Esteban2010Tmproved}, geodesy \cite{exertier2019time}, astronomy \cite{Schediwy2019The}, deep space exploration \cite{Xiao2018PhotonicsbasedWD} and communication \cite{Mahmood2019Time, Sliwczynski2020Fiber}, and has garnered significant attention. In recent years, quantum clock synchronization (QCS) has come into the limelight. QCS is a clock synchronization technology based on quantum entanglement.It obtains a high-precision clock difference by calculating the time difference between entangled photon pairs, can avoid symmetric delay attacks,
\Authorfootnote
 \noindent
and has the characteristics of high precision and high security. The successfully verified QCS schemes mainly include two-way QCS \cite{hou2019fiber, hong2023quantum}, HOM interferometer based QCS \cite{quan2019simulation}, and round-trip QCS \cite{tang2023demonstration, lee2022absolute}.
 At present, the majority of the work focuses on the time synchronization between the server side and a single user side. With the development of quantum networks, an increasing number of users have begun to join. Hence, a QCS network capable of serving multiple users is of paramount importance. In 2023, Tang proposed a QCS network based on a quantum entanglement network. This network employs a PPLN entangled light source based on spontaneous parametric down-conversion (SPDC) and realizes QCS through the round-trip scheme. The channel resources used by $n$ users amount to $2n$. It achieved a time deviation (TDEV) as low as \SI{426}{\femto\second} within an average time of \SI{4000}{\second} \cite{tang2023demonstration}. Nevertheless, this scheme contains a large quantity of free-space devices, making the system integration challenging. It is susceptible to the environment and is not readily applicable in large quantities to the scenarios demanded in reality.
 
 \par Spontaneous four-wave mixing (SFWM), as another major effect for preparing entangled light sources, can occur on numerous on-chip material platforms such as Si 
  \cite{wang2024progress}, $\mathrm{Si_{3}N_{4}}$ 
  \cite{wen2023polarization}, AlGaAs 
  \cite{mahmudlu2021algaas}, etc. Among them, the silicon on insulator (SOI) platform has a vast number of maturely designed devices, such as filters \cite{guan2015ultra}, polarization beam splitters \cite{Li:20}, arrayed waveguide grating(AWG) 
  \cite{li2017design}, etc. Furthermore, the silicon-based waveguide can be well integrated with the standard CMOS process and exhibits good compatibility \cite{Steglich_2021}, which simplifies system integration. SFWM is a nonlinear effect caused by third-order optical nonlinearity, described by the coefficient $\chi_{3}$. It is mainly manifested as the system absorbing two pump photons to generate a pair of signal-idler photons. When two pump photons with the same (different) frequencies are considered degenerate (non-degenerate) SFWM \cite{Signorini:18}. According to the wavelength conversion characteristics of SFWM \cite{Kultavewuti:16, 8082195}, the generated entangled photon pairs can form a quantum entanglement distribution network through the wavelength division multiplexing strategy \cite{ PhysRevApplied.18.024059}.
 This method can improve the transmission capacity and fully utilize broadband entangled photons to realize applications such as quantum computing, quantum communication, QCS, etc \cite{feng2020progress, li2017chip, djordjevic2020global, tang2023demonstration}. 
 
\par In this paper, we propose a QCS network scheme based on a silicon-chip dual-pumped SFWM \cite{guo2018experimentally} photon source. Utilizing the multi-channel entanglement generated by dual pumped, the entangled photons are distributed to remote users through the wavelength division multiplexing strategy to construct an entanglement distribution network, and the round-trip QCS is adopted to realize a QCS network that can serve multiple users.  The channel resources used by $n$ users in this network are \(\lceil\frac{4n}{3}\rceil\), which is reduced by approximately 30\% compared to the previous network. Additionally, in the case of using a single entangled photon source, the number of users can be linearly expanded by simply adding relevant devices and optical fibers. 
We conducted a proof-of-principle experiment with a server and three users (Alice, Bob, and Charlie) for 11.1 hours. Alice and Charlie are \SI{10}{\kilo\metre} away from the server while Bob is \SI{25}{\kilo\metre} away from the server. 
The lowest TDEVs between the server and each user (Alice, Bob, and Charlie) are \SI{1.57}{\pico\second}, \SI{0.82}{\pico\second} and \SI{2.57}{\pico\second} at the average time of \SI{8000}{\second}, \SI{8000}{\second} and \SI{800}{\second} respectively.
\section{QCS network scheme}\label{sec:2}
\subsection{Dual pumped SFWM}
This scheme utilizes dual-pumped SFWM to generate time-energy entangled photon sources.
Pumped photons at angular frequencies $\omega_{p}$ and $\omega_{q}$ simultaneously enter the silicon-based waveguide. 
When photons of $\omega_{p}$ interact with photons of $\omega_{p}$, degenerate SFWM occurs, resulting in entangled photon pairs with $\omega_{s}$ and $\omega_{i}^{1}$  (eq. (\ref{pp})).   
When photons of $\omega_{p}$ interact with photons of $\omega_{q}$, non-degenerate SFWM occurs to produce entangled photon pairs with $\omega_{s}$  and  $\omega_{i}^{2}$ (eq. (\ref{pq})).  
When photons of $\omega_{q}$ interact with photons of $\omega_{q}$, degenerate SFWM occurs to produce  entangled photon pairs with $\omega_{s}$ and $\omega_{i}^{3}$ (eq. (\ref{qq})).
Here $\omega_{p}$ and $\omega_{q}$ are narrow band sources, while $\omega_{s}$, $\omega_{i}^{1}$, $\omega_{i}^{2}$ and $\omega_{i}^{3}$ are broad band photons.
\begin{figure}[H]
\centering
\includegraphics[width=\linewidth]{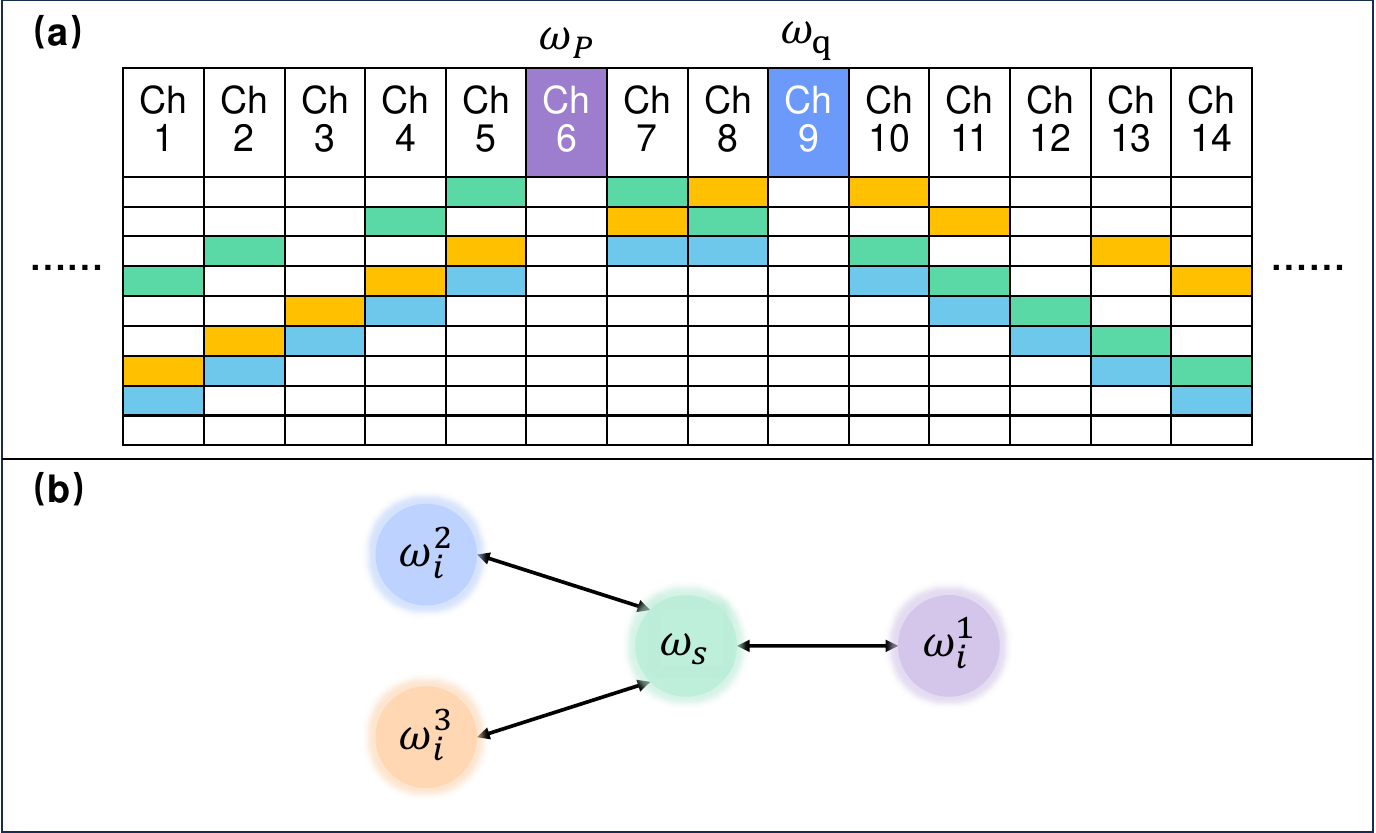}
\caption{(a) Dual-pumped SFWM generates multi-channel entangled photon pairs, among which Ch6 and Ch9 are the channels for $\omega_{p}$ and $\omega_{q}$. The color blocks of the same color in each row correspond to a pair of entangled photons. The green (golden)  blocks represent multiple pairs generated by degenerate SFWM with Ch6 (Ch9) as the pump channel. The cyan blocks represent multiple pairs generated by non-degenerate  SFWM with Ch6 and Ch9 as the pump channels.  (b) The entanglement photons will be generated among one signal channel and three idler channels.}
\label{Dual_pumped_SFWM}
\end{figure}
With a dense wavelength division multiplexing (DWDM), we could divide the generated photons into multiple channels, as shown in Figure \ref{Dual_pumped_SFWM} (a), where we take 14 channels as an example, Ch6 and Ch9 are the pump channels.
 \begin{figure*}[ht]
\centering{\includegraphics[width=0.7\linewidth]{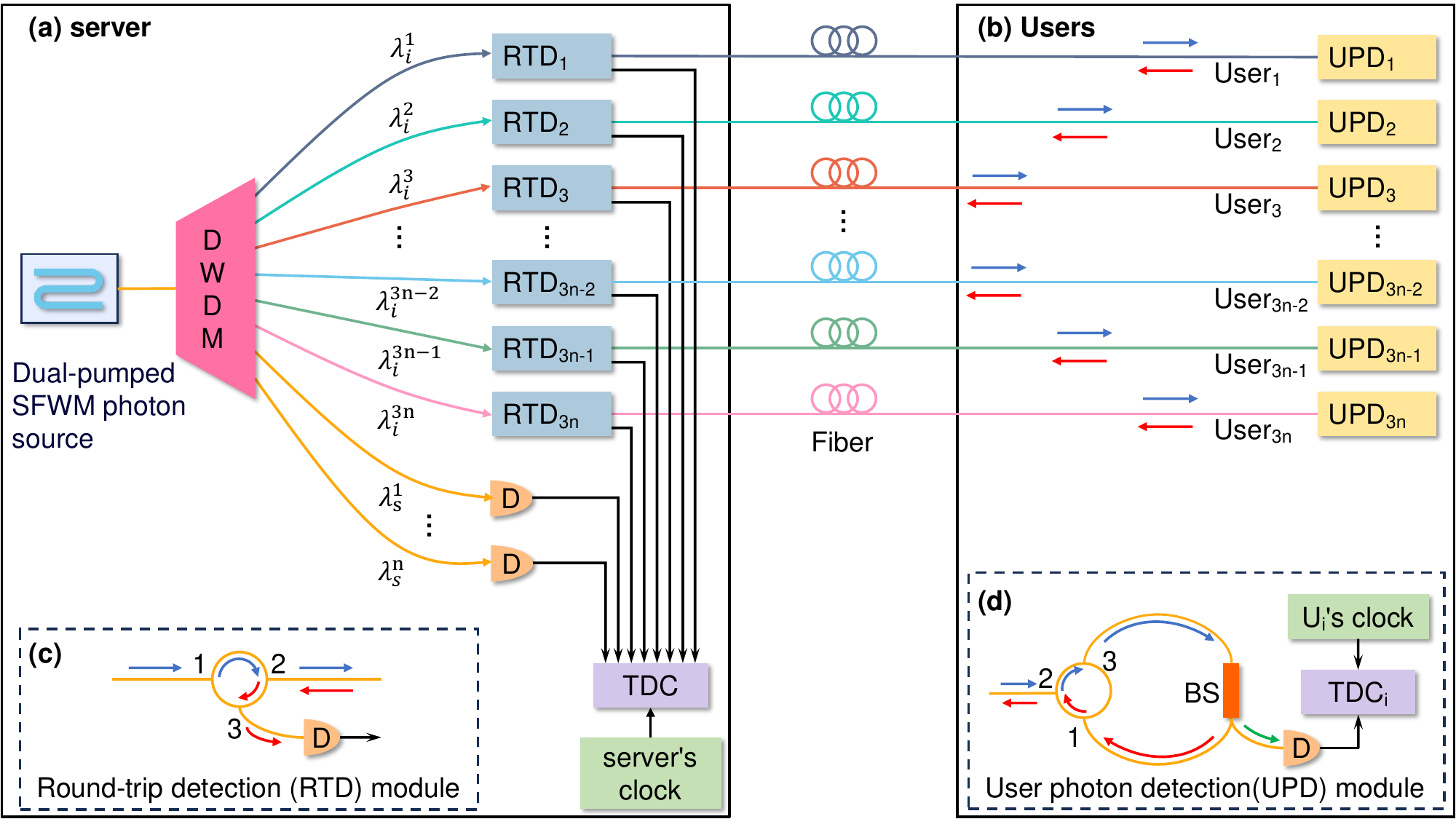}}
  \caption{Schematic diagram of a multi-user round-trip QCS scheme based on dual-pumped SFWM. (a) The entangled photon pairs are demultiplexed by DWDM. The idler photons are then transmitted to the user after passing through a round-trip detection (RTD) module. The signal photons are transmitted to a detector for detection. (b) After the photons transmitted from the server to the user enter a user photon detection (UPD) module, the detector detects a portion of the photons, while the other portion returns to the server. (c) RTD module: the input photons pass through an optical circulator (OC) from port 1 to port 2 and the round-trip photons are detected by a detector.  (d) UPD module: the input photons pass through an OC (from port 2 to port 3) and a beam splitter (BS), with a portion being detected directly and another portion being sent back to the server through port 1 to port 2 of the OC. The server’s and user’s detectors are connected to their respective time-to-digital converters (TDC), which are synchronized by their local clocks.}
  \label{schematic diagram}
  \end{figure*}
\begin{align}
  \omega_{p}+\omega_{p}=\omega_{s} +\omega_{i}^{1}
  \label{pp}
  \\ \omega_{p}+\omega_{q}=\omega_{s} +\omega_{i}^{2}
  \label{pq}
 \\ \omega_{q}+\omega_{q}=\omega_{s} +\omega_{i}^{3}
 \label{qq}
\end{align}
 The green (golden)  blocks represent multiple pairs generated by degenerate SFWM with Ch6 (Ch9) as the pump channel. The cyan blocks represent multiple pairs generated by non-degenerate  SFWM with Ch6 and Ch9 as the pump channels. In each row, color blocks of the same color correspond to a pair of entangled photons, and the covered channels can expand regularly to the left and right directions.  For each channel (excluding Ch3 and Ch12), it can be paired with three other channels, as shown in Figure \ref{Dual_pumped_SFWM} (b). Based on the requirements of networking and the complexity of the wiring, multiple entanglements between one signal channel and three idler channels as shown in Figure \ref{Dual_pumped_SFWM} (b) can be used to construct the QCS network. 
 \subsection{QCS network scheme}
 The proposed QCS network is illustrated as Figure \ref{schematic diagram}. The DWDM demultiplexes the entangled photon pairs generated by the dual-pumped SFWM photon source into photons with wavelengths of $\lambda _{i}^{j}$ (idler) and $\lambda _{s}^{k}$ (signal), $j\in[1,3n], k\in [1,n], n \in N^{*}$. The idler photons with the wavelength of $\lambda _{i}^{j}$ pass through the user round-trip detection module (RTD) and then enter the user photon detection (UPD) module after traveling through a section of optical fiber, while the signal photons with the wavelength of $\lambda _{s}^{k}$ are directly detected. The network complexity of this scheme is \(O(n)\), and the channel resources used by \(n\) users are \(\lceil\frac{4n}{3}\rceil\).  Here, we take the clock synchronization process between $\mathrm{User_{1}}$ and the server as an example for a detailed explanation. 
 The signal photons with the wavelength of $\lambda _{s}^{1}$ are separated from the DWDM and directly detected at the server. The idler photons with the wavelength of $\lambda _{i}^{1}$ are sent into $\mathrm{RTD_{1}}$, pass through optical circulate (OC) (from port 1 to port 2), and a section of optical fiber to reach $\mathrm{UPD_{1}}$, and are transmitted to BS from port 3 of OC. The photons are divided into two parts by beam splitter (BS). One part is directly detected, and one part enters from port 1 of OC, exits from port 2, and returns to the $\mathrm{RTD_{1}}$ of the server. The returned photons are transmitted from port 3 of OC and detected. The outputs of all detectors are connected to the time-to-digital converter (TDC) synchronized with the clock of the server. 
 \par Since the entangled photon pairs generated by SFWM are time-correlated \cite{zhang2019generation}, the time differences between the signal photons and the idler photons can be determined by using the second-order correlation function $G^{(2)}$($\tau$) \cite{glauber1963quantum}. The time when the signal photons are detected is recorded as $t_{1}$, the time when the idler photons are detected at $\mathrm{UPD_{1}}$ is recorded as $t_{2}$, and the time when the idler photons return to $\mathrm{RTD_{1}}$ and are detected is recorded as $t_{3}$. 
 The time difference $\tau _{s-u}$ between $t_{1}$ and $t_{2}$ consists of $\Delta t_{s-u}$ (the time needed for the photons to reach $\mathrm{User_{1}}$ from the server) and the clock difference $\Delta t_{1}$, which can be expressed as
   \begin{equation}
     \tau _{s-u} = t_{2} -t_{1} = \Delta t_{s-u} + \Delta t_{1}
     \label{eq1}
   \end{equation}
 The time difference $\tau_{s-u-s}$ between $t_1$ and $t_3$ represents the 
  \begin{figure*}[ht]
 \centering\includegraphics[width=0.7\linewidth]{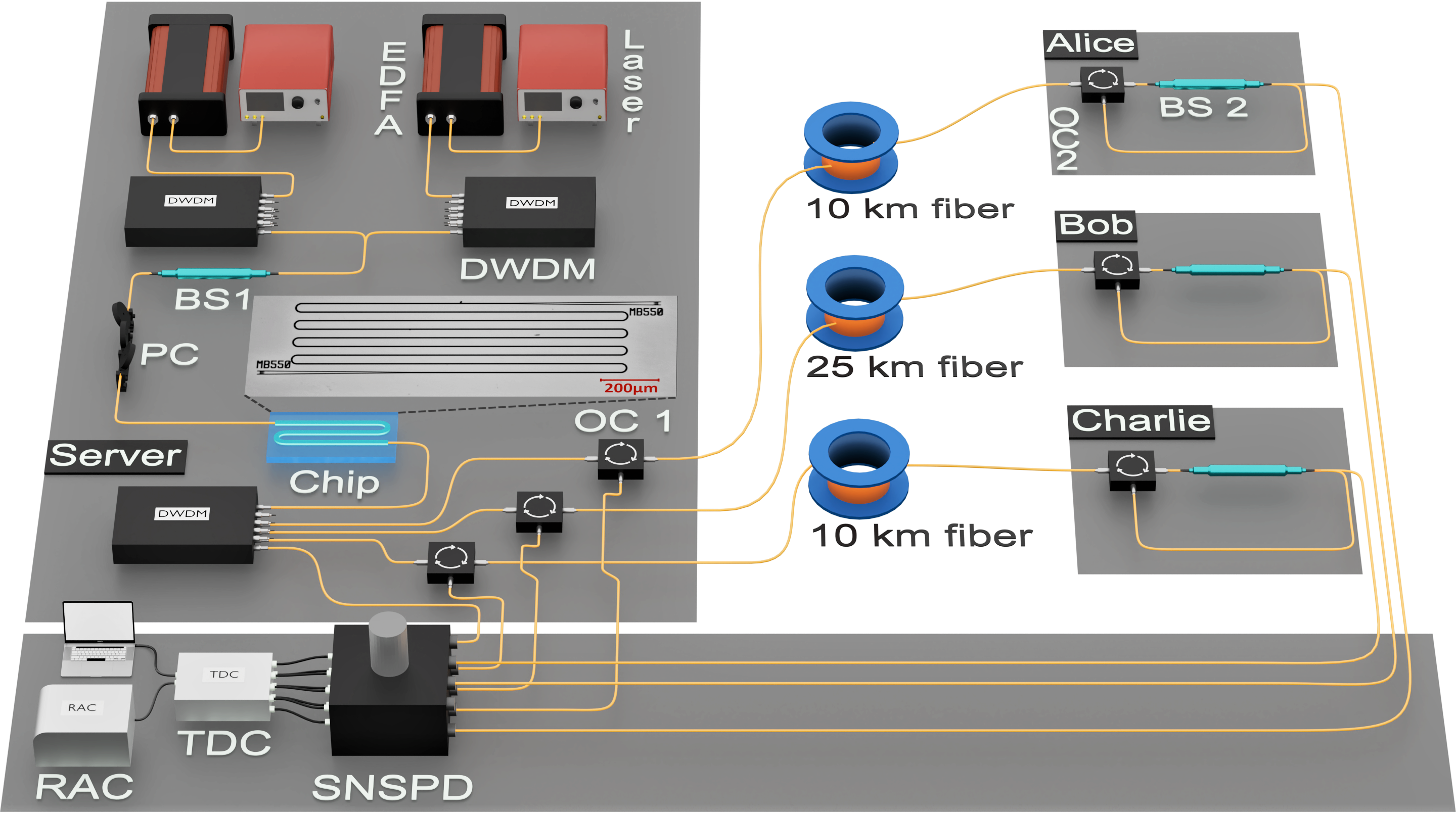}
   \caption{Experimental setup of round-trip multiple users (one server and three users) QCS based on dual-pumped SFWM. EDFA: erbium-doped fiber amplifier, DWDM: dense wavelength division multiplexer, BS: beam splitter, PC: polarization controller, OC: optical circulator, SNSPD: superconducting nanowire single-photon detectors, RAC: rubidium atomic clock.}
   \label{device}
 \end{figure*}
 whole round-trip propagation time $\Delta t_{s-u-s}$ of the photons between the server and $\mathrm{User_{1}}$, which can be expressed as 
   \begin{equation}
     \tau _{s-u-s} = t_{3} -t_{1} = \Delta t_{s-u-s} 
     \label{eq2}
   \end{equation}
   Since the photons have the same round-trip propagation time between the server and $\mathrm{User_{1}}$, it can be derived that the clock difference is expressed as
 \begin{equation}
   \Delta t_{1}=\tau _{s-u}-\frac{\tau _{s-u-s}}{2}
   \label{eqdt}
 \end{equation}
 \par The clock difference $\Delta t_{a}$ between other users and the server can also be calculated using the method described above, where $a\in [1,3n], n \in N^{*}$. The same method can be applied to calculate the clock difference for other users as well. The clock difference between any two users $\mathrm{User_{a}}$ and $\mathrm{User_{b}}$ can be expressed as
 \begin{equation}
   \Delta t _{a-b}=\Delta t _{a}-\Delta t _{b}
   \label{eq4}
 \end{equation}
 Where $a, b\in [1, 3n], n \in N^{*}$.
\section{Experimental results}\label{sec:3} 
\subsection{Experimental setup}
In the experiment, we set up a server and three users  (Alice, Bob, and Charlie) to demonstrate the proposed QCS network, shown in Figure \ref{device}. 
The pump source is generated by two tunable lasers with center wavelengths of \SI{1555.75}{\nano\meter} and \SI{1544.53}{\nano\meter}, corresponding to International Telecommunication Union (ITU, \SI{100}{GHz}) channels C27 and C41, respectively. The pump photons are amplified by two erbium-doped fiber amplifiers (EDFA). The amplified spontaneous emission (ASE) noise and pump sideband noise are filtered out using a DWDM. Then, 
$\mathrm{BS_{1}}$  combines the two pump beams into one beam. The polarization state of the pump photons is controlled by a polarization controller (PC) and coupled into a silicon-based waveguide sample through a photonic crystal grating coupler. 
After passing through the silicon-based waveguide, the residual pump photons and the signal/idler photons are simultaneously coupled into another DWDM, which is used to filter out the residual pump photons and perform demultiplexing of the generated signal-idler photon pairs. 
The signal photons correspond to ITU C33, centered at \SI{1550.92}{\nano\meter}. The idler  photons correspond to C21 (centered at \SI{1560.61}{\nano\meter}), C35 (centered at \SI{1549.32}{\nano\meter}), and C49 (centered at \SI{1538.19}{\nano\meter}), respectively. 
The signal photons are detected directly by the superconducting nanowire single-photon detector (SNSPD) after exiting from the DWDM and are time-stamped by the TDC. The TDC's clock is synchronized with the location's clock.
Alice and Charlie are both \SI{10}{\kilo\meter} away from the server through C49 and C21, respectively. Bob is \SI{25}{\kilo\meter} away from the server through C35.
We take the time synchronization between Alice and the server as an example to describe the experiment.
Photons from C49 undergo DWDM demultiplexing and an $\mathrm{OC_{1}}$ and then travel over \SI{10}{\kilo\meter} of fiber to $\mathrm{OC_{2}}$ and $\mathrm{BS_{2}}$, where it is split into two parts. One part is transmitted to Alice for detection. 
The other part will return to the server by re-enter  $\mathrm{OC_{2}}$ and the \SI{10}{\kilo\meter} fiber. The returned photons arrive at the server from port 3 of $\mathrm{OC_{1}}$ and are detected by the SNSPD. Due to the limited experimental equipment, all the photons were detected by one SNSPD. The outputs of SNSPD are connected to a TDC. Furthermore, by synchronizing the clock of the TDC with the rubidium atomic clock (RAC), the clock drift and jitter are reduced, thereby improving the accuracy of time measurement.  
\subsection{Data analysis}
\begin{figure}[H]
   \centering{\includegraphics[width=1\linewidth]{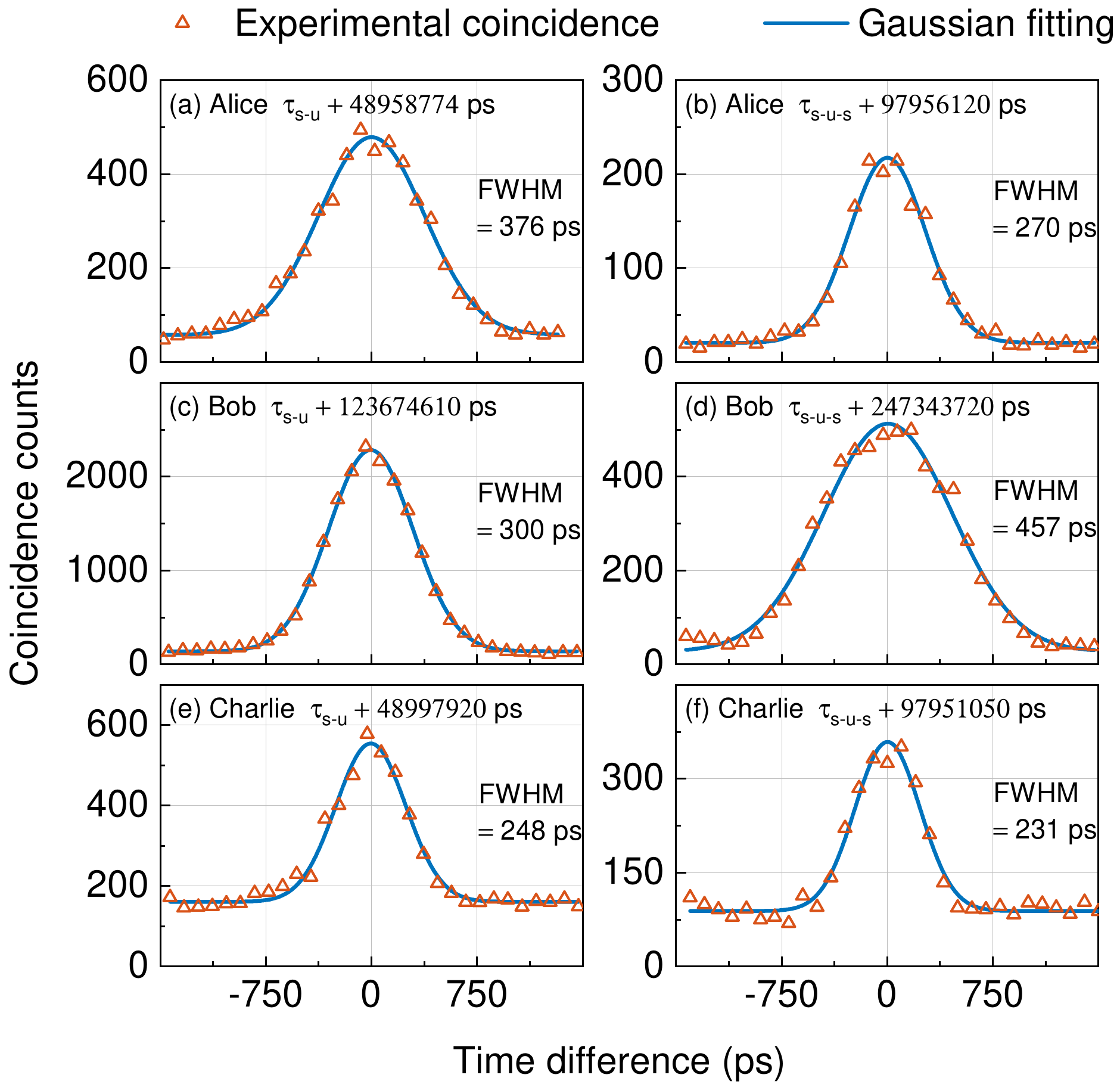}}
   \caption{The coincidence count histograms of $G^{(2)}(\tau_{s-u})$ and $G^{(2)}(\tau_{s-u-s})$ in the first \SI{20}{\second}  experiment.  (a) and (b) show the coincidence histograms $G^{(2)}(\tau_{s-u})$ and $G^{(2)}(\tau_{s-u-s})$ of Alice, while (c) and (d) as well as (e) and (f) show the same for Bob and Charlie, respectively.}{}
   \label{gas}
\end{figure}

\begin{figure}[H]
 \centering{\includegraphics[width=\linewidth]{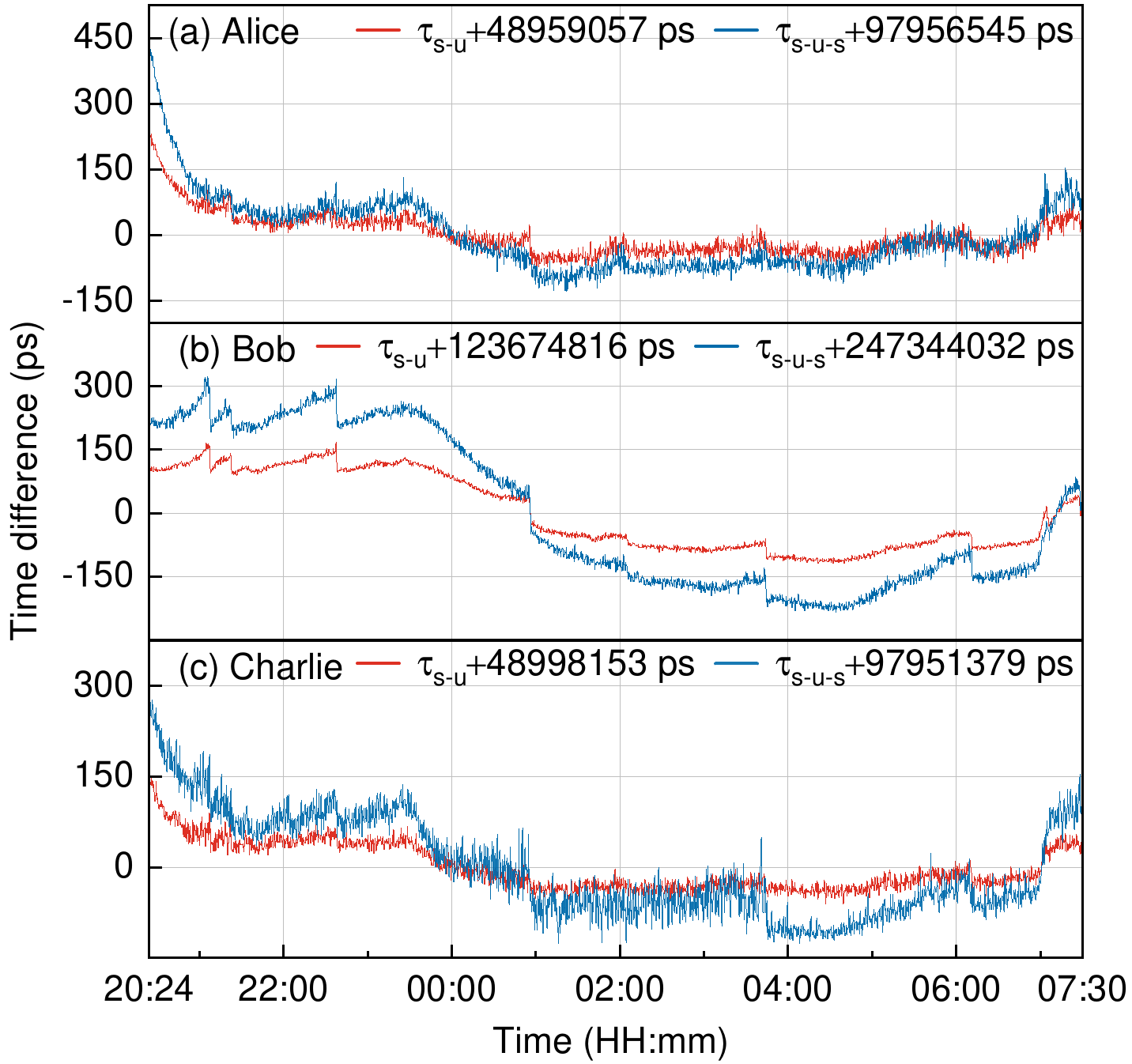}}
   \caption{Measured time differences between each user and the server at intervals of \SI{20}{\second}.}
   \label{offset}
 \end{figure}

\begin{figure}[H]
 \centering{\includegraphics[width=\linewidth]{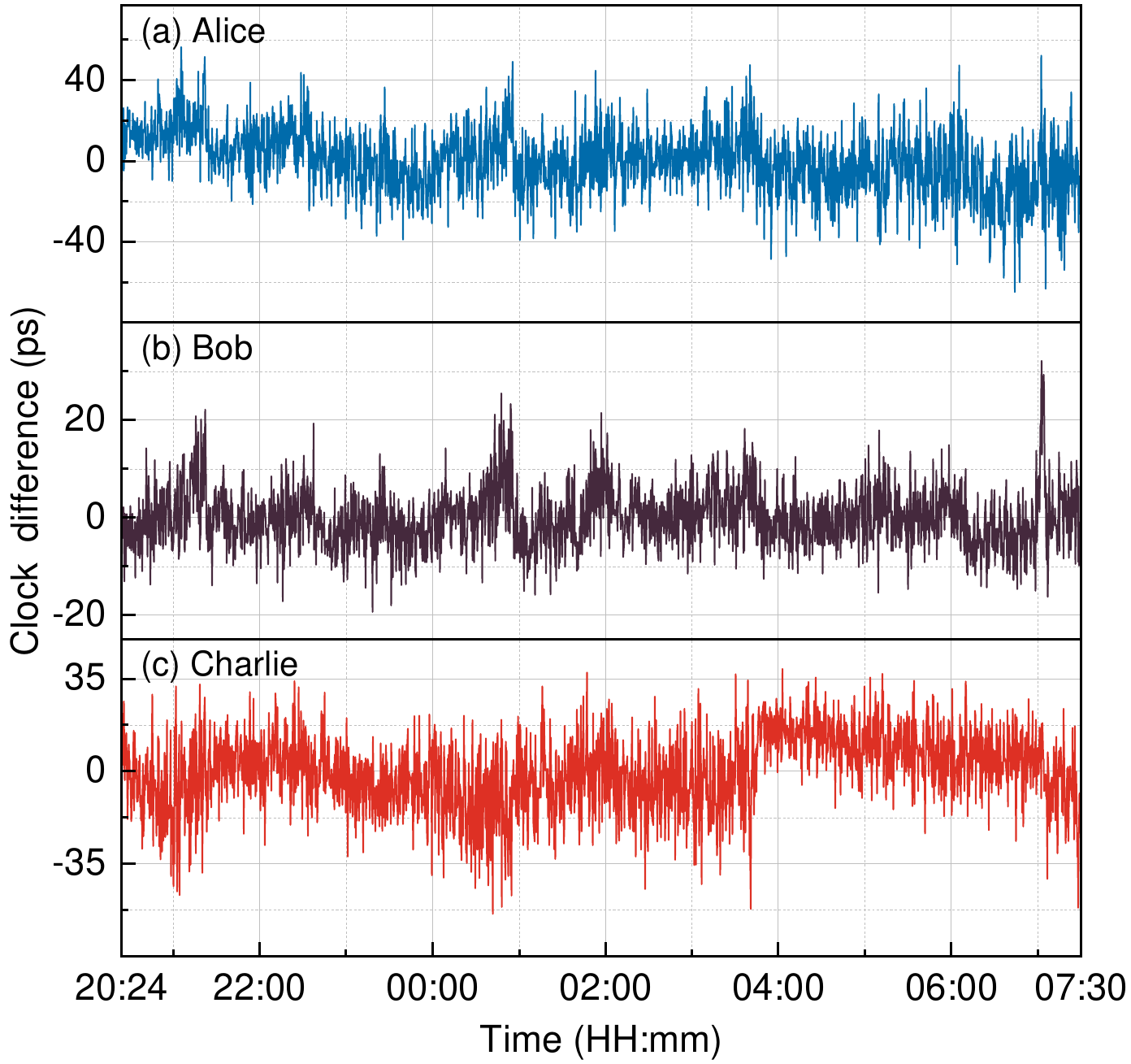}}
   \caption{The clock differences between each user and the server measured in the experiment.}
   \label{dt}
 \end{figure}

 During the experiment, the time differences $\tau_{s-u}$ and $\tau_{s-u-s}$ were calculated at intervals of \SI{20}{\second} using the eqs. (\ref{eq1}) and (\ref{eq2}) to ensure that the system is real-time. The Gaussian fit results for the first \SI{20}{\second} are shown in Figure \ref{gas}. Due to the presence of group velocity dispersion in optical fibers, the full width at half maximum (FWHM) of Bob's $G^{(2)}(\tau_{s-u-s})$ coincidence histogram is greater than the FWHM of $G^{(2)}(\tau_{s-u})$. However, Alice and Charlie exhibit the opposite phenomenon. This is because the noise levels of each channel are different, and SNSPD working under different counting rates will have varying degrees of time drift \cite{mueller2023time}. Therefore, the extreme point and FWHM of the coincidence histogram will change, resulting in the FWHM of the coincidence histogram observed in Alice and Charlie not conforming to the law of fiber dispersion.
The time differences $\tau _{s-u}$ and $\tau _{s-u-s}$ between the server and the users
 \noindent
over consecutive 11.1 hours are shown in Figure \ref{offset}. 
$\tau _{s-u}$ and $\tau _{s-u-s}$ for each user are calculated from the fitting results. The $\tau _{s-u}$ ($\tau _{s-u-s}$) expectations between the server and each user (Alice, Bob, and Charlie) are $\SI{-48959057}{ps}$ ($\SI{-97956545}{ps}$), $\SI{-123674816}{ps}$ ($\SI{-247344032}{ps}$), and $\SI{-48998153}{ps}$ ($\SI{-97951379}{ps}$), respectively.  
According to Figure \ref{gas}, Bob's coincidences-to-accidentals ratio (CAR) is significantly higher than Alice and Charlie's, resulting in 
 \noindent
smaller fluctuations in the time difference between Bob and the server in Figure \ref{offset} compared to other users. Furthermore, due to the instability of the chip coupling platform itself, we manually adjust the coupling efficiency every period during the experiment to ensure the stability of the number of photon pairs generated. Since the time drift of the SNSPD depends on the count rate of each channel \cite{mueller2023time}, this behavior will cause sudden changes in the count rate of each channel, leading to sudden changes in the time differences $\tau _{s-u}$ and $\tau _{s-u-s}$ (the extreme points of the coincidence histogram). 
As shown in Figure \ref{offset}, the variation in $\tau_{s-u-s}$ for each user is approximately twice that of the variation in $\tau_{s-u}$, therefore, when calculating the clock difference, these variations tend to cancel each other out. The clock difference $\Delta t$ can be calculated according to eq. (\ref{eqdt}), as shown in Figure \ref{dt}. The standard deviations of the clock difference between the server and each user (Alice, Bob, and Charlie) are \SI{16.2006}{\pico\second}, \SI{6.2810}{\pico\second}, and \SI{14.6764}{\pico\second}, respectively. The reason for $\Delta t$ not being zero can be attributed to the transmission time asymmetry of the various system components (including BS, OC, DWDM, PC, etc.) as well as the time drift of SNSPD.
 \noindent
\par
\begin{figure}[H]
   \centering
   \includegraphics[width=\linewidth]{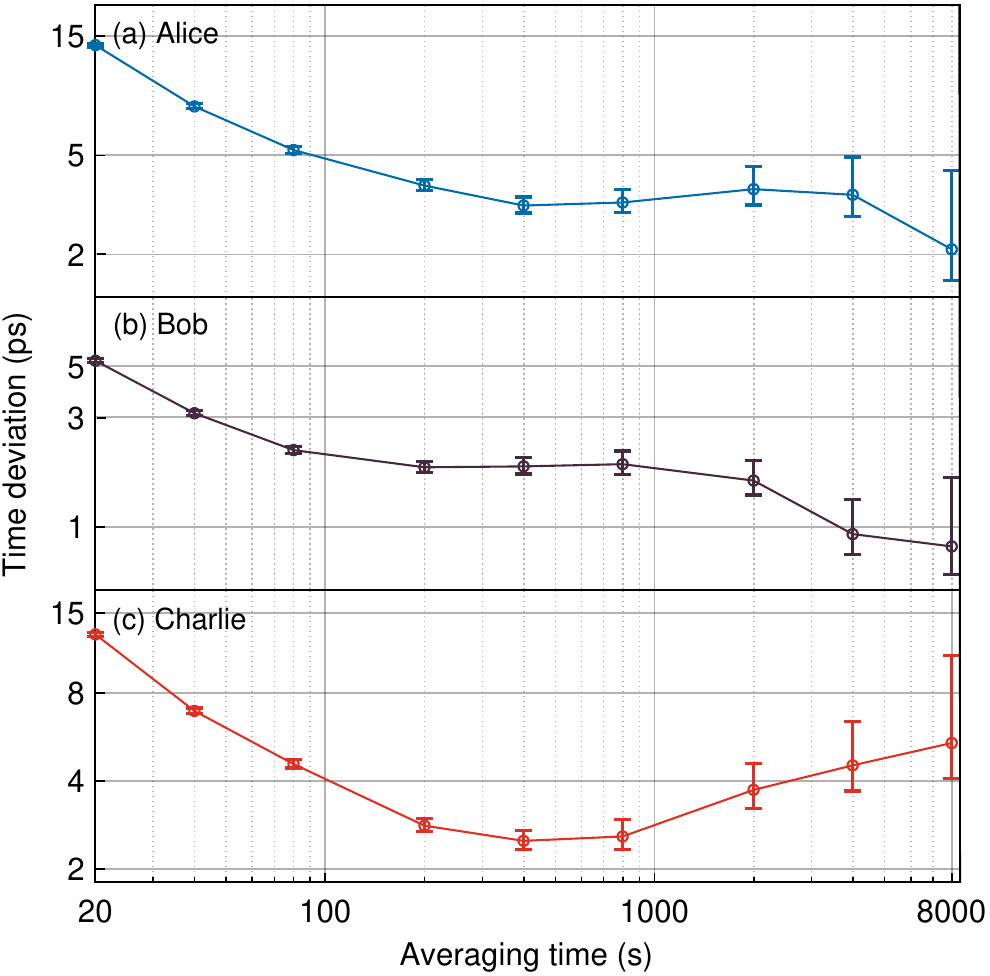}
   \caption{TDEV of clock differences measured by each user in the experiment.}
   \label{TDEV}  
\end{figure}

 Figure \ref{TDEV} illustrates the stability of the clock difference between the server and each user in terms of TDEV. 
The TDEVs between the server and each user (Alice, Bob, and Charlie) are \SI{3.14}{\pico\second}, \SI{1.84}{\pico\second}, and  \SI{2.49}{\pico\second} at an average time of \SI{400}{\second}, respectively. The lowest TDEVs between the server and each user (Alice, Bob, and Charlie) are \SI{1.57}{\pico\second}, \SI{0.82}{\pico\second} and \SI{2.57}{\pico\second} at the average time of \SI{8000}{\second}, \SI{8000}{\second} and \SI{800}{\second} respectively. 
\vspace{1em}
As mentioned, Bob's TDEV is still better than the other users.The experimental results demonstrate the feasibility and effectiveness of the multi-user QCS scheme based on a silicon chip dual-pumped SFWM for multi-user high-precision QCS.
\section{Conclusion and Discussion}\label{sec:4}
In conclusion, we put forward a QCS network scheme based on a silicon-chip dual-pumped SFWM photon source, distributing photons to remote users by means of the wavelength division multiplexing strategy. This scheme employs the entanglement between one signal channel and three idler channels, considerably reducing the demand for channel resources. The channel resources needed for $n$ users is \(\lceil\frac{4n}{3}\rceil\), which is approximately 30\% lower than that of the previous scheme. To verify the feasibility of the scheme, we carried out a multi-user clock synchronization experiment involving a server and three users, lasting for 11.1 hours. 
Two of the users were \SI{10}{\kilo\meter} away from the server, and one user was \SI{25}{\kilo\meter} away from the server. The TDEVs between the server and each user (Alice, Bob, and Charlie) are \SI{3.14}{\pico\second}, \SI{1.84}{\pico\second}, and  \SI{2.49}{\pico\second} at an average time of \SI{400}{\second}, respectively. The lowest TDEVs between the server and each user (Alice, Bob, and Charlie) are \SI{1.57}{\pico\second}, \SI{0.82}{\pico\second} and \SI{2.57}{\pico\second} at the average time of \SI{8000}{\second}, \SI{8000}{\second} and \SI{800}{\second} respectively. 
The final results of TDEV were mainly influenced by the chip coupling efficiency, the filtering performance of the DWDM, the detection efficiency of the SNSPD, and the jitter of the TDC. 
However, under these influences, the Time Deviation (TDEV) between Alice and the server in the experiment still reached the sub-picosecond level at around \SI{4000}{\second}, indicating that this scheme has great potential to reduce the TDEV to the femtosecond level.  And by using more entanglements between one signal channel and three idler channels, it can simply serve more users. In the future, multiple on-chip devices can be integrated onto a single chip to lower system costs and enhance stability, enabling QCS to serve more users in more complex environments.
\Acknowledgements{This research is funded by the Guangxi Science and Technology Major Special Project, China under Grants AA23073013; National Key R$\&$D Program of China (2022YFF0706005); National Natural Science Foundation of China (12272407, 62275269, 62275271, 62305387); State Key Laboratory of High Performance Computing, National University of Defense Technology (202201-12); The Natural Science Foundation of Hunan Province (2023JJ40683, 2022JJ40552); The science and technology innovation Program of Hunan Province (2023RC3003).}

\InterestConflict{The authors declare that they have no conflict of interest.}





\end{multicols}
\end{document}